\documentstyle[12pt,aaspp]{article}
\begin{document}
\def\wc{\hangindent=4em \hangafter=1 \noindent}

\title{THE MASSES OF NEARBY DWARFS AND BROWN DWARFS WITH THE HST
\footnote{Submitted for publication to Acta Astronomica, Vol. 46}}
\author{Bohdan Paczy\'nski}

\affil{Princeton University Observatory, Princeton, NJ 08544--1001}
\affil{E-mail: bp@astro.princeton.edu}

\begin{abstract}
The known nearby stars, moving in front of the background of distant
stars and galaxies, create 'weak gravitational lensing' variations
in their positions.  These variations may be measurable with the HST,
and they may allow a direct mass determination for the nearby stars.
The cross section for the HST measurable astrometric effect
is much larger than the cross section for the photometric effect
which is measurable from the ground.  The mass determination will
be easier for the fainter nearby lenses which will be discovered in
future searches of faint high proper motion stars.
\end{abstract}


\vskip 0.5cm

Any massive object moving in front of distant sources makes
their positions vary by deflecting their light rays, the phenomenon
known as gravitational lensing (cf. Paczy\'nski 1996 for references
and the derivation of all formulae).  The effect is presented in Figure 1,
where the trajectories of double images of distant sources with
respect to the lens are shown with curved solid lines, and the
Einstein ring is shown with the dashed line.  One image is always
formed on the outside of the ring, while the second image is always 
formed inside the ring.

The possibility
of measuring gravitational lensing effects astrometrically was analyzed
by Hog et al. (1995), Miyamoto and Yoshi (1995), and Gould (1996).
They discussed
lensing events caused by objects located at a distance of many kiloparsecs,
so the two images were expected to be separated by about a milli arcsecond,
and the lens itself was assumed to be too dark to be visible.  Such
lensing events may be resolved, or the light centroid displacement
may be measured with future space based optical interferometers.
Miralda-Escud\'e (1996) proposed to look at relatively bright nearby stars
with ground based near-infrared interferometers, and to measure the
astrometric displacement of the background stars caused by gravitational
lensing  by the nearby stars.  Both approaches require new technology
to be developed, and it is not clear how long we have to wait for that.
The purpose of this paper is to point out that a similar astrometric
effect may be measurable now with the HST for a few dozen of the nearest
dwarf stars, leading to the determination of their masses.
This is an astrometric analogue of the photometric project described
before (Paczy\'nski 1995).

\begin{figure}[t]
\vspace{12cm}
\includegraphics{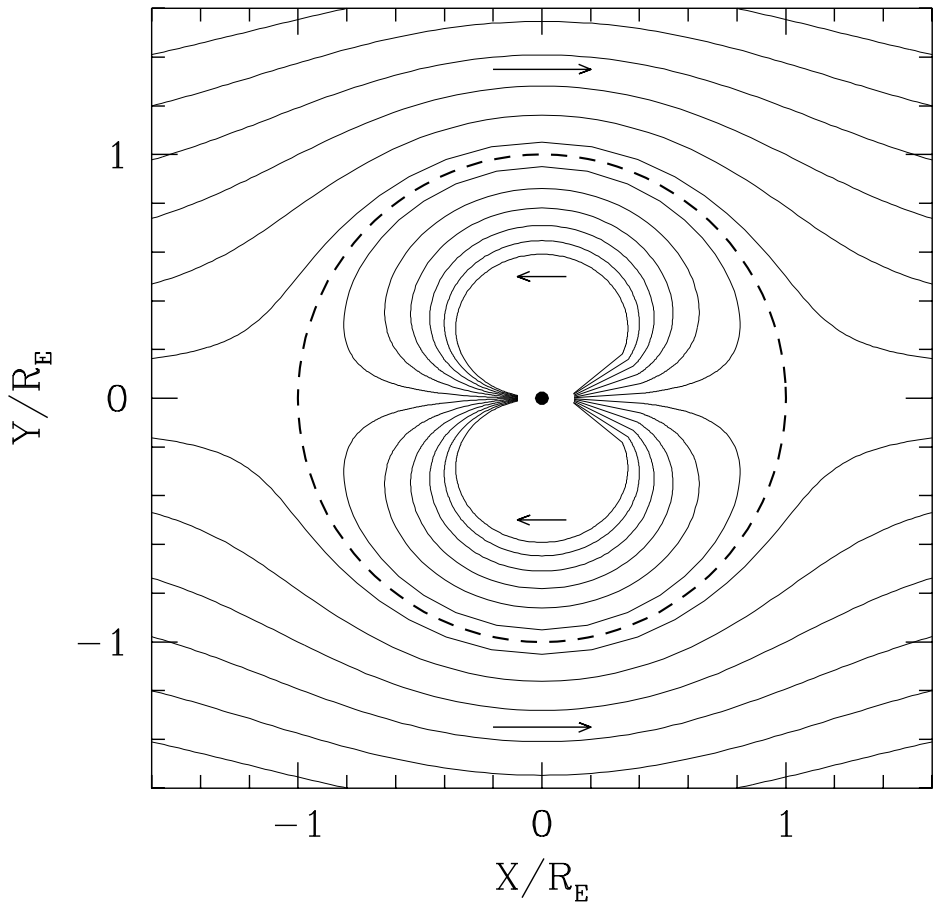}
\caption{\small
The geometry of gravitational lensing of many sources
by a single lens is presented in projection onto the sky.
The trajectories of images of distant sources with respect to 
the gravitational lens are shown with solid lines, the
Einstein ring is shown with the dashed line, and the lens is
shown with a large dot at the center.  One image is always
formed on the outside of the Einstein ring, while the second image 
is always formed inside the ring.
}
\end{figure}

The angular radius of the Einstein ring is given as
$$
\varphi _E = 
\left[ {4GM \over c^2 } ~ { D_s - D_d \over D_d D_s } \right] ^{1/2} =
0.''023 \times 
\left( { M \over 0.2 ~ M_{\odot}} ~ { 3 ~ pc \over d_{\pi} } \right) ^{1/2} ,
\eqno(1)
$$
where $ M $ is the lens mass and $ d_{\pi} $ is the parallax distance 
to the lens as measured with respect to the source:
$$
{ 1 \over d_{\pi} } = { 1 \over D_d } - { 1 \over D_s }  .
\eqno(2)
$$
We consider here a few dozen dwarf stars located
at a distance of $ D_d \sim 3 ~ pc $, with the sources being typically
at $ D_s > 1 ~ kpc $.

If a source is located at an angular distance $ \varphi _s $ from
a lens, then the two images are located at angular distances
$ \varphi _{\pm} $ from the lens, according to
$$
\varphi _{\pm} = 0.5 ~ \varphi _E \left[ u 
\pm 0.5 \left( u^2 + 4 \right) ^{1/2} \right],
\hskip 1.0cm u \equiv { \varphi _s \over \varphi _E }  .
\eqno(3)
$$
For a large value of the dimensionless impact parameter $ u $ we have
approximate relations
$$
\varphi _+ \approx \varphi _s + { \varphi _E^2 \over \varphi _s } , 
\hskip 1.0cm \varphi _- \approx - { \varphi _E^2 \over \varphi _s } .
\eqno(4)
$$
The magnification of the brightness of the two images is given as
$$
A_{\pm} = { u^2 +2 \over 2u \left( u^2 + 4 \right) ^{1/2} } \pm 0.5 ,
\hskip 1.0cm  
A = A_+ + A_- = { u^2 +2 \over u \left( u^2 + 4 \right) ^{1/2} },
\eqno(5)
$$
i.e.
$$
A_+ = 1 + A_- \approx 1 + \left( { \varphi _E \over \varphi _s } \right) ^4 , 
\hskip 2.0cm {\rm for} \hskip 0.5cm 
u = { \varphi _s \over \varphi _E } \gg 1 .
\eqno(6)
$$
Note, that for large values of the impact parameter $ u $ the
increase in apparent brightness is proportional to $ u^{-4} $, while
the displacement in the image locations is proportional to
$ u^{-1} $.  In other words, the effect of gravitational lensing
on the image positions falls off with the impact parameter much less
less rapidly than the effect on the apparent brightness, as pointed
out by Miralda-Escud\'e (1996), and references therein.
Therefore, if the brighter of the two images and the lens
are resolved then the cross section for an astrometric effect is much 
larger than the cross section for a photometric effect.

Let us consider as an example the Barnard's star (cf. van de Kamp 1971).
It has a parallax of $ 0.''552 $ and a proper motion of $ 10.''31 $
per year.  If we adopt the mass of $ \sim 0.2 ~ M_{\odot} $ we obtain for 
its Einstein radius 
$$
\varphi _{E,Barnard} = 
0.''030 \times 
\left( { M_{Barnard} \over 0.2 ~ M_{\odot}} \right) ^{1/2} ,
\eqno(7)
$$
The HST can resolve two point like images at a separation of 
$ \varphi _{HST} \approx 0.''1 $.  The displacement due
to gravitational lensing by the Barnard's star at the impact
parameter of $ \varphi _+ $ is expected to be
$$
\Delta \varphi _+ = \varphi _+ - \varphi _s
= { \varphi _{E,Barnard}^2 \over \varphi _+ }
\approx 0.''009 ~ \left( { M_{Barnard} \over 0.2 ~ M_{\odot} } \right)
~ \left( { 0.''1 \over \varphi _+ } \right) , 
\eqno(8)
$$

A displacement as small as $ 0.''002 $ should be accurately measurable with 
the HST, which gives $ \sim 1'' $ as the total geometrical cross section for
astrometrically measurable gravitational lensing.   Combined with the high
proper motion this gives the total area of $ \sim 10 \times (1'')^2 $
covered by the Barnard's star in one year, which is much more than
$ \sim 0.02 \times (1'')^2 $ covered by a typical high proper motion 
object considered by Paczy\'nski (1995).  The good news is that the
cross section for an astrometic effect is much larger than a cross
section for a photometric effect.  The bad news is that it may be very 
difficult for the 
HST to measure the positions of the faint background stars or galaxies 
located within $ 0.''1 $ of the bright Barnard's star, which has
$ V \approx 9.5 $.  However, with the spectral type of M5 the Barnard's
star may be sufficiently faint in the ultraviolet to make the relative
astrometry feasible.  Note, that the mass of the lensing star as given
with the equation (8) is proportional to the astrometric displacement
$ \Delta \varphi _+ $, i.e. the measurement of the displacement is equivalent
to a direct measurement of the lens mass.

There are a few dozen nearby stars in the van den Kamp's (1971) list,
some of them considerably fainter than Barnard's star, like Wolf 359,
with a parallax of $ 0.''431 $, a proper motion of $ 4.''71 $ per year, 
an apparent magnitude $ V \approx 13.5 $, and a spectral type M8e.
The estimate of the Einstein radius for Wolf 359 is
$$
\varphi _{E,Wolf ~ 359} = 
0.''023 \times 
\left( { M_{Wolf ~ 359} \over 0.15 ~ M_{\odot}} \right) ^{1/2} ,
\eqno(9)
$$
It is possible, and even likely, that large proper
motion stars even fainter than Wolf 359 will be discovered (Alard 1996),
and will become prime brown dwarf candidates.
Accurate measurement of their masses with the HST will be crucial
for the determination of their masses, and therefore their nature.
Excellent field brown dwarf candidates have recently reported by
Hawkins and Jones (1996) in their relatively low proper motion study, 
which revealed four objects at $ \sim 30 ~ pc $.

Let us consider now a case when with a proper choice of filters the lensing
star can be extinguished, and all that is measurable is an unresolved
pair of the two lensed images, with the locations and magnifications
given with the eqs. (3) and (5).  The displacement of light centroid
can be calculated as
$$
\Delta \varphi \equiv \varphi - \varphi _s =
{ ( \varphi _+ - \varphi _s) A_+ + ( \varphi _- - \varphi _s) A_- \over A }
= { u \over u^2 +2 } ~ \varphi _E  ,
\eqno(10)
$$
which reaches the maximum value of $ 2^{-3/2} \varphi _E \approx 0.354 ~
\varphi _E $ for $ u \equiv \varphi _s / \varphi _E = 2^{1/2} \approx 1.414 $.
These displacements are large and easily measurable even from the ground
with the existing instruments, provided the light from the nearby lens can
somehow be extinguished, or the lens is simply very faint.

\begin{figure}[t]
\vspace{12cm}
\includegraphics{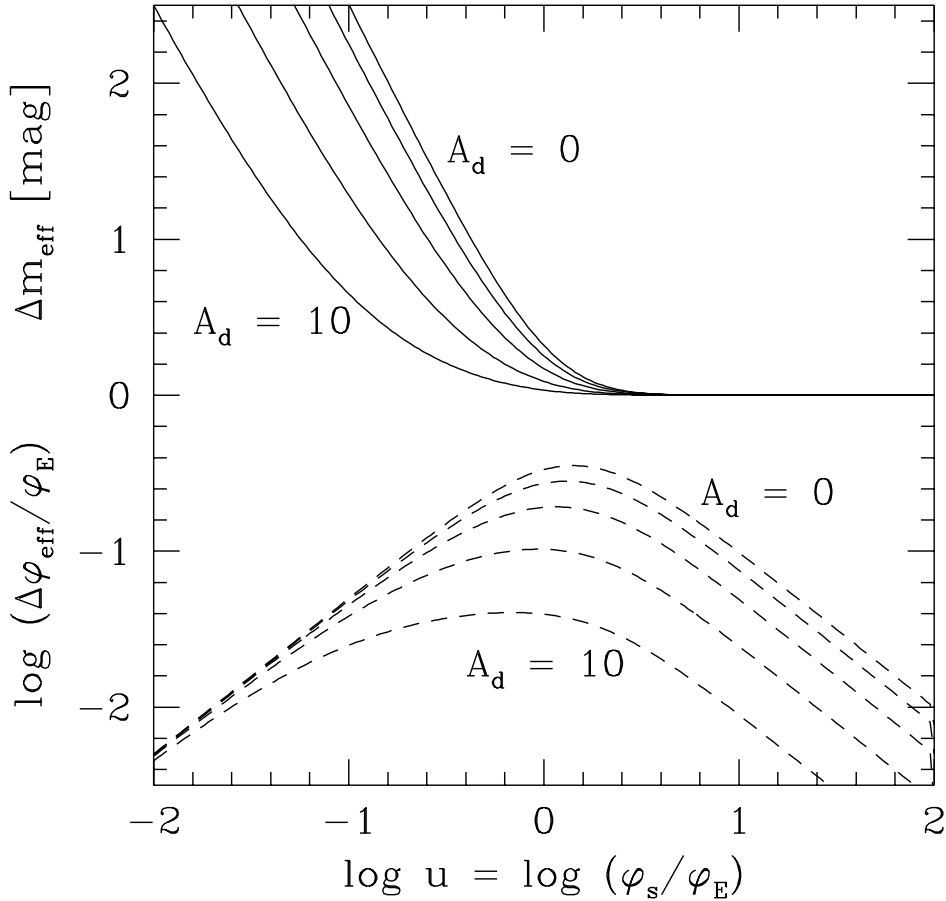}
\caption{\small
The variations of the photometric ($ \Delta m_{eff} $) and astrometric
($ \Delta \varphi _{eff} / \varphi _E $) effects of gravitational lensing 
are shown as a function of dimensionless impact parameter
($ u = \varphi _s / \varphi _E $) with solid and dashed lines,
respectively.  The five lines correspond to five values of the
ratio: $ A_d \equiv F_{lens}/F_{source} = 0, ~ 0.3, ~, 1, ~ 3, ~ 10 $,
respectively.
}
\end{figure}

In some cases, perhaps even in most cases, the lens and the unresolved 
double image of the lensed star will combine to form a single composite 
image.  Therefore, it is useful to compare the amplitude of the photometric 
and astrometric effects expected for such blended microlensing events.  Let 
the observed flux from the lens be $ A_d $ times the flux from the source.  
The amplitude of the astrometric effect, i.e. the difference in the position
between the light centroid of the {\it lens plus double image system} and the 
centroid of the {\it lens plus source system} in the absence of microlensing
is given as
$$
\Delta \varphi _{eff} = \Delta \varphi ~ { A(u) \over A(u) + A_d } =
{ u \over u^2 +2 } ~ { A(u) \over A(u) +A_d } ~\varphi _E  ,
\hskip 1.0cm A_d \equiv { F_{lens} \over F_{source} } ,
\eqno(11)
$$
where $ A(u) $ is given with the eq. (5).  The photometric effect,
i.e. the brightness of the composite image in units of the combined
brightness of the lens and the unlensed source is given as
$$
A_{eff} = { A(u) + A_d \over 1 + A_d } , \hskip 1.0cm
\Delta m_{eff} \equiv 2.5 \log A_{eff} .
\eqno(12)
$$

The dependence of the astrometric effect,
$ \Delta \varphi _{eff} / \varphi _E $, and the photometric effect,
$ \Delta m_{eff} $, on the impact parameter $ u $ is shown in Figure 2
for five values of the relative lens brightness: $ A_d = 0, ~ 0.3, ~
1, ~ 3, ~ 10 $.  $ A_d = 0 $ corresponds to the lens which
is invisible, and $ A_d = 10 $ corresponds to the lens which is 10 times
brighter than the source.  Naturally, the fainter the lens the stronger
the astrometric and photometric effects.  Note, that the photometric
effect becomes rapidly negligible for the impact parameter $ u > 1 $,
i.e. $ \varphi _s > \varphi _E $, while the astrometric effect may
be measurable even for much larger values of the impact parameter.
For example, for the lens as bright as the source ($ A_d = 1 $)
and $ \varphi _s = 3 \varphi _E $ the effective displacement of
the light centroid is $ \Delta \varphi _{eff} = 0.14 \varphi _E $.
This corresponds to $ \sim 0.''004 $ for the Barnard's star.  This is
a small displacement, but it should be easily measurable with the HST,
and perhaps even from the ground, from a site with an excellent seeing.

The cross section for astrometric effect of gravitational microlensing
is larger than the photometric effect, but both require the lens to
be relatively dim with respect to the source.  As the two stars are
likely to have different spectral types their brightness ratio
can be optimized with a proper selection of filters.  This task may be 
very difficult with the currently known nearby stars, as these are
relatively bright.  Future searches of faint high proper motion
stars and/or brown dwarfs (Alard 1996, Hawkins and Jones 1996) will
lead to a discovery of objects much more 
suitable for microlensing based mass determination.  Note, that the
faint objects will also be the most interesting, being candidates for
brown dwarfs and the faintest (and hence the oldest) degenerate dwarfs.
Cool objects like M dwarfs and brown dwarfs have strong molecular
bands which should make it relatively easy to select filters in which
these objects will appear very faint, i.e. with a very small value
of $ A_d $ (cf. eq. 11).

Let us compare the relative merits of the project proposed here with
those proposed by Miralda-Escud\'e (1996) and Paczy\'nski (1995).
The project proposed here has an advantage over that proposed by
Paczy\'nski (1995) in that it can be carried out over the whole
sky, not only in the Milky Way, thanks to the much larger cross section
for the astrometric effects of gravitational lensing as compared
to the cross section for the photometric effects.  The weakness is
the requirement of the HST resolving power as opposed to a 1-meter class 
ground based photometric telescope.  Still, if some objects are very rare,
like hypothetical brown dwarfs, the ability to conduct the search over
the whole sky may turn out to be essential.  The current project
may work best for the very faint high proper motion objects, while
the project proposed by Miralda-Escud\'e (1996) requires relatively 
bright objects as lenses, so the two projects are complementary.

{\bf Acknowledgements.} It is a great pleasure to thank Dr. P. Madau,
Dr. J. Miralda-Escud\'e, and Dr. K. Sahu for stimulating discussions.
This project was supported with the NSF grants 
AST 92-16494 and  AST 93-13620.

\vskip 0.5cm
\centerline{REFERENCES}
\vskip 0.5cm

\wc{Alard, C. 1996, private communication.  \hfill}

\wc{Gould, A. 1996, {\it Publ. Astron. Soc. Pacific}, in press = 
astro-ph/9604014.  \hfill}

\wc{Hawkins, M. R. S. and Jones, H. 1996, presented at the {\it Royal
Astronomical Society's National Astronomy Meeting} in Liverpool (reported
in {\it Science News}, {\bf 149}, 345).  \hfill}

\wc{Hog, E., Novikov, I. D., and Polnarev, A. G. 1995, {\it Astron. and
Astrophys.}, {\bf 294}, 287.  \hfill}

\wc{Miralda-Escud\'e, J. 1996, preprint astro-ph/9605138.  \hfill}

\wc{Miyamoto, M., and Yoshi, Y.  1995, {\it Astron. J.}, {\bf 110}, 1427,
\hfill}

\wc{Paczy\'nski, B. 1995, {\it Acta Astronomica}, {\bf 45}, 345.  \hfill}

\wc{Paczy\'nski, B. 1996, {\it Ann. Rev. Astron. Astrophys.}, {\bf 34},
in press.  \hfill}

\wc{van de Kamp, P. 1971, {\it Ann. Rev. Astron. Astrophys.}, {\bf 9}, 103.
\hfill}

\end{document}